\documentclass[a4paper,11pt]{article}
\pdfoutput=1 

\usepackage{jheppub} 

\usepackage[T1]{fontenc} 

\title{\boldmath Conformal Complex Singlet Extension of the Standard Model: Scenario for Dark Matter and a Second Higgs Boson}

\author[a,1]{Zhi-Wei Wang,\note{Corresponding author.}}
\author[a]{T.G.~Steele}
\author[b]{T.~Hanif}
\author[c]{R.B.~Mann}

\affiliation[a]{Department of Physics and
Engineering Physics, University of Saskatchewan,\\116 Science Place, Saskatoon, SK,
S7N 5E2, Canada}
\affiliation[b]{Department of Theoretical Physics, University of Dhaka,\\Dhaka-1000, Bangladesh}
\affiliation[c]{Department of Physics, University of Waterloo,\\Waterloo, ON, N2L 3G1, Canada}

\emailAdd{zhw283@mail.usask.ca}
\emailAdd{tom.steele@usask.ca}

\emailAdd{rbmann@uwaterloo.ca}
\emailAdd{thanif@du.ac.bd}

\abstract{We consider a conformal complex singlet extension of the Standard Model with a Higgs portal interaction. The global $U(1)$ symmetry of the complex singlet can be either broken or unbroken and we study each scenario. In the unbroken case, the global $U(1)$ symmetry protects the complex singlet from decaying,  leading to an ideal cold dark matter candidate with approximately 100 GeV mass along with a significant proportion of thermal relic dark matter abundance. In the broken case, we have developed a renormalization-scale optimization technique to significantly narrow the parameter space and in some situations, provide unique predictions for all the model's couplings and masses.  
We have found there exists a second Higgs boson with a mass of approximately $550\,\rm{GeV}$ that mixes with the known $125\,\rm{GeV}$ Higgs with a large mixing angle $\sin\theta\approx 0.47$ consistent with current experimental limits.
The imaginary part of the complex singlet in the broken case could provide axion dark matter for a wide range of models. Upon including interactions of the complex scalar with an additional vector-like fermion,  we explore the possibility of a diphoton excess in both the unbroken and the broken cases. In the unbroken case, the model can provide a natural explanation for diphoton excess if extra terms are introduced providing extra contributions to the singlet mass. In the broken case, we find a set of coupling solutions that yield a second Higgs boson of mass $720\,\rm{GeV}$ and an $830\,\rm{GeV}$ extra vector-like fermion $F$, which is able to address the $750\,\rm{GeV}$ LHC diphoton excess. We also provide criteria to determine the symmetry breaking pattern in both the Higgs and hidden sectors.}

\begin{document} 
\maketitle
\flushbottom

\section{Introduction}
\label{sec:intro}

The Standard Model (SM) of particle physics is incomplete since it does not provide an explanation for dark matter. Amongst the numerous ways to go beyond the SM, Higgs portal models \cite{zee,Cline:ab} 
are conceptually appealing because they provide a link between Higgs hunting in collider experiments and dark matter direct detection experiments \cite{Djouadi:2012zc}. Complex singlet extensions with global $U(1)$ symmetry yield rich phenomenological properties, such as a second Higgs particle mixed with the ordinary Higgs particle along with WIMP dark matter candidates \cite{Barger:2008jx,Gonderinger:2012rd,Costa:2014qga}. The global $U(1)$ symmetry also provides a foundation for further model-building \cite{Basso:2010yz,Iso:2012jn,Khoze:2014xha}, in particular interactions with an extra vector-like fermion \cite{Franceschini:2015kwy,Knapen:2015dap,Zhang:2015uuo} that may explain the LHC diphoton excesses \cite{ATLAS,CMS}.


Versions of hidden sector extensions with classical conformal symmetry are particularly interesting since they can address the hierarchy and naturalness problems \cite{Susskind,Bardeen,'tHooft:1979bh} associated with the conventional electroweak symmetry breaking mechanism.  Conformal symmetry as a custodial symmetry protects the Higgs mass from large UV contributions, which addresses the naturalness problem \cite{Bardeen,'tHooft:1979bh}.  In this case, the conformal symmetry can only be softly broken and needs to be restored sufficiently quickly \cite{Tavares:2013dga}. In addition, if the electroweak symmetry breaking is realized by the Coleman-Weinberg (CW) mechanism within conformal models, a natural scale hierarchy is generated through the dimensional transmutation similar to the QCD case \cite{Weinberg:1978ym,Hill:2014mqa}. In these models, there exists two main interpretations for the origin of electroweak (EW) symmetry breaking that are usually associated with different ranges of the couplings. In the first, radiative symmetry breaking (RSB, or the CW mechanism \cite{Coleman:1973jx}) in the hidden (dark) sector gets communicated to the Higgs sector. This triggers EW symmetry breaking via the Higgs portal interaction \cite{a}, which is normally negative (see e.g.~Ref.~\cite{Altmannshofer:2014vra}). Alternatively, RSB could occur in the SM Higgs sector first and then be communicated to the hidden sector. In this second interpretation, a reasonably large Higgs quartic coupling is usually required to balance the large top quark Yukawa coupling and a positive Higgs portal interaction is permitted \cite{Steele:2013fka,Elias:2003zm,Wang}. 


We consider here two main scenarios depending on whether or not the global $U(1)$ symmetry is spontaneously broken by a vacuum expectation value of the hidden-sector complex field. For the broken $U(1)$ case, we extend  and improve the optimization method proposed in \cite{Casas:1994us} to multiple scalar fields to accommodate the addition of a complex-singlet vacuum expectation value. Generalizing this method to incorporate RSB, we find that in addition to the SM Higgs particle, there exists a second Higgs boson with a $554\,\rm{GeV}$ mass. We also explore including extra vector-like fermions and find a set of viable solutions where the mass of the second Higgs boson  increases to around $720\,\rm{GeV}$, which is able to address the $750\,\rm{GeV}$ diphoton anomaly observed at the LHC \cite{ATLAS,CMS}. This also leads to an axion dark matter candidate whose properties depend on detailed model building. This  improved optimization method  depends on local properties rather than global searchers, and therefore has very strong predictive power, affording dynamical generation of all the parameters in the model. In the unbroken case, we find a large Higgs self-coupling perturbative regime similar to Refs.~\cite{Elias:2003zm,Wang}, along with a scalar dark matter candidate  that provides a significant proportion of  dark matter abundance.  We have also explored the possibility of diphoton excess in the unbroken case and find that a natural explanation of the diphoton excess can be provided only if extra terms are introduced in the singlet (hidden) sector to increase the singlet mass.

\section{Model}

The complex singlet extension of the SM  with an extra vector-like fermion $F$ has the Lagrangian \cite{Barger:2008jx,Knapen:2015dap,Zhang:2015uuo}:
\begin{equation}
\begin{split}
L&=\frac{1}{2}\partial_\mu H^{\dagger}\partial^\mu H+\frac{1}{2}\partial_\mu S^{\dagger}\partial^\mu S-\lambda_2\left|S\right|^2 H^{\dagger}H-\lambda_3\left|S\right|^4\\
&-\lambda_1\left(H^{\dagger}H\right)^2 +i\bar{F}\gamma^{\mu}D_{\mu}F-\left(y_S\bar{F_L}F_RS+\rm{h.c.}\right)
\end{split}
\end{equation}
where  $F$ transforms as $\left(R_C,R_W\right)_{Y_F}$, $H$ is the (complex doublet) Higgs field and $S$ is the complex singlet field.  Here we assume the diphoton excess is realized through the process $gg\rightarrow S\rightarrow \gamma\gamma$ where $g$ represents the gluon and two vector-like fermion $F$ loops are required at both the production and decay process. The LHC so far has not provided any hints in other channels, leading to strong upper bounds on other decay channels of the $S$-resonance \cite{Franceschini:2015kwy}. It is therefore crucial that $S$ has no direct interactions with SM fields except via the Higgs portal interaction proportional to $\lambda_2$, which prevents the large decay channels of the resonances to the SM particles as well as preventing large suppression of the diphoton excesses by the large decay width of $S$ to top quark $\Gamma_{S\rightarrow tt}$ \cite{Knapen:2015dap}. The above Lagrangian obeys a global $U(1)$ symmetry for $S$. This symmetry may either be unbroken ($\langle S\rangle=0$) or broken ($\langle S\rangle\neq0$), and we consider each case in turn.  Note that the diphoton excess will also be dependent on the above symmetry breaking pattern. In the broken case, the singlet $S$ will mix with the SM Higgs, which opens other decay channels of $S$ through the mixing. If the mixing angle is not small enough, the upper bounds  of $S$ decay to other SM particles \cite{Franceschini:2015kwy} will be violated.

\section{Unbroken Phase}

For the unbroken case, $S$ decay is protected by the $U\left(1\right)$ global symmetry, making it an ideal cold dark matter candidate when $y_S$ is set to zero. In addition, this case may also provide a natural explanation for the diphoton excess when $y_S$ is turned on since $S$ will not mix with the Higgs and the decay channels of $S$ to other SM particles are greatly suppressed.  Our analysis builds upon the Gildener-Weinberg method \cite{gildener} that generalizes the CW technique \cite{Coleman:1973jx} to incorporate multiple scalar fields. 
Letting $H=\frac{1}{\sqrt{2}}\left(\phi_1+i\phi_2, \phi_3+i\phi_4\right)$, $S=\frac{1}{\sqrt{2}}\left(\varphi_1+i\varphi_2\right)$ and defining $\phi^2=\sum_{i}\phi_i^2$ and $\varphi^2=\sum_{i}\varphi_i^2$, we obtain leading-logarithm expression for the effective potential
\cite{Steele:2013fka}
\begin{equation}
V_{LL}=\frac{1}{4}\lambda _1 \phi^4+\frac{1}{4}\lambda _2\phi^2\varphi^2+\frac{1}{4}\lambda _3\varphi^4+BL+CL^2+DL^3+EL^4+\ldots\label{VLL}
\end{equation}
where $L\equiv\log\left(\frac{\phi^2+\varphi^2}{\mu^2}\right)$. 
The quantities $B, C, D, E$ are the functions of $\left(\lambda_1,\lambda_2,\lambda_3,g_t,\phi,\varphi\right)$ which are dimension-4 combinations of $\phi^2$ and $\varphi^2$ as required by 
 symmetry and
contain leading-logarithm ($LL$) combinations of couplings $\left(\lambda_1^{\alpha}\lambda_2^{\beta}\lambda_3^{\gamma}g_t^{2\delta}\right)L^p$ 
where $g_t$ is the top Yukawa coupling and $p-\left(\alpha+\beta+\gamma+\delta\right)=1$.  The coefficients $B, C, D, E$ are determined by Renormalization Group (RG) equation
\begin{equation}
\left(\mu\frac{\partial}{\partial\mu}+\beta_{g_t}\frac{\partial}{\partial g_t}+\sum_{i=1}^3\beta_i\frac{\partial}{\partial \lambda_i}+
\gamma_\phi \phi\frac{\partial}{\partial \phi}\right)V_{LL}=0\label{rg equation}
\end{equation}
where the  one loop RG functions $\beta_i,\beta_{g_t}$ and anomalous dimensions $\gamma_\phi$ are given by \cite{Zhang:2015uuo,EliasMiro:2012ay}
\begin{equation}
\begin{split}
\beta_1&=\frac{1}{16\pi^2}\left(24\lambda_1^2+\lambda_2^2-6g_t^4+12\lambda_1g_t^2\right)\\
\beta_2&=\frac{1}{16\pi^2}\lambda_2\left(8\lambda_3+12\lambda_1+4\lambda_2+6g_t^2+2R_CR_Wy_S^2\right)\\
\beta_3&=\frac{1}{16\pi^2}\left(2\lambda_2^2+20\lambda_3^2+ 2R_CR_Wy_S^2\left(2\lambda_3-y_S^2\right) \right)\\
\beta_{g_t}&=\frac{1}{16\pi^2}\left(\frac{9}{2}g_t^3\right)\quad,\quad\gamma_\phi=\frac{1}{64\pi^2}\left(12g_t^2\right)\,.\label{RG functions}
\end{split}
\end{equation}
and the anomalous dimension for the singlet field $\gamma_\varphi=0$ at one loop order.
Truncation of the effective potential at $LL$ order requires counter terms  corresponding to those in the  Lagrangian
\begin{equation}
V_{eff}=V_{LL}+K_1\phi^4+K_2\phi^2\varphi^2+K_3\varphi^4 
\label{counter1}
\end{equation}
where $K_i$ are functions of the couplings.

Defining $\rho^2=\phi^2+\varphi^2$ \cite{gildener}, the three renormalization conditions in the CW (or Jackiw) scheme \cite{Coleman:1973jx,Jackiw:1974cv} used to determine $K_i$  
are conveniently expressed as \cite{Japanese}
\begin{equation}
\frac{d^4V_{eff}}{d\rho^4}\bigg|_{\rho=\mu}=\frac{d^4V_{tree}}{d\rho^4}\bigg|_{\rho=\mu}\label{CW Condition}
\end{equation}
where $V_{tree}$ is the tree level effective potential.

To determine the couplings, we need to employ the vacuum expectation value (VEV) 
conditions, which provide constraints for the minimum of the vacuum 
\begin{equation}
\frac{dV_{eff}}{d\phi}\bigg|_{\phi=v\atop \varphi=v_1}=0\quad,\quad \frac{dV_{eff}}{d\varphi}\bigg|_{\phi=v\atop \varphi=v_1}=0\,.\label{constraint2}
\end{equation}
where $v$ is identified with the electroweak scale $v=246.2\,\rm{GeV}$. In the unbroken case ($v_1=0$), the above singlet VEV condition is trivial since it identically vanishes whereas this is not true in the broken case ($v_1\neq0$). In the unbroken case, we also identify the renormalization scale $\mu$ with the electroweak scale $\mu=v=246.2$ to eliminate the higher-logarithmic terms. The mass generated for the Higgs doublet $M_H$ and singlet $M_S$ are only dependent on the second-order terms in the effective potential and can be determined from the eigenvalues of the mass matrix $M$
\begin{equation}
M=\left(
\begin{array}{cc}
\frac{dV_{eff}^2}{d\phi^2} & \frac{dV_{eff}^2}{d\phi d\varphi} \\
\frac{dV_{eff}^2}{d\varphi d\phi} & \frac{dV_{eff}^2}{d\varphi^2}
\end{array}
\right)\Bigg|_{\phi=v\atop \varphi=v_1}\label{mass}
\end{equation}
where in the unbroken case, the off-diagonal terms are zero and we obtain $M_{H}^2=\frac{dV_{eff}^2}{d\phi^2}\big|_{\phi=v\atop \varphi=v_1}$, $M_{S}^2=\frac{dV_{eff}^2}{d\varphi^2}\big|_{\phi=v\atop \varphi=v_1}$. Note that we have implicitly used the result that the effective potential kinetic term renormalization constant is unity in the CW scheme \cite{Coleman:1973jx,Jackiw:1974cv}.

Consider first the unbroken symmetry case with $y_S=0$ (no contributions from $F$).  Eq.~\eqref{constraint2} only contains one non-trivial constraint, and hence it is not possible to constrain all the couplings.   We find that altering the singlet self-interaction coupling within the range $0<\lambda_3<1$ affects the physical dark matter mass predictions by less than $2.4\%$. We therefore set $\lambda_3=0$, corresponding to the case of 
weakly self-interacting dark matter, commenting on $\lambda_3 \neq 0$ as appropriate. The Higgs portal interaction $\lambda_2$ is then the only input parameter;  it will be strongly constrained by dark matter abundance and direct detection experiments XENON100 \cite{Aprile:2012nq} and LUX \cite{Akerib:2013tjd}. The Higgs mass prediction in this case is consistent with our previous findings and converges to approximately $125\,\rm{GeV}$ when higher loop contributions are included \cite{Steele:2013fka,Wang}.

\begin{figure}[htb]
\centering
\includegraphics[width=0.76\columnwidth]{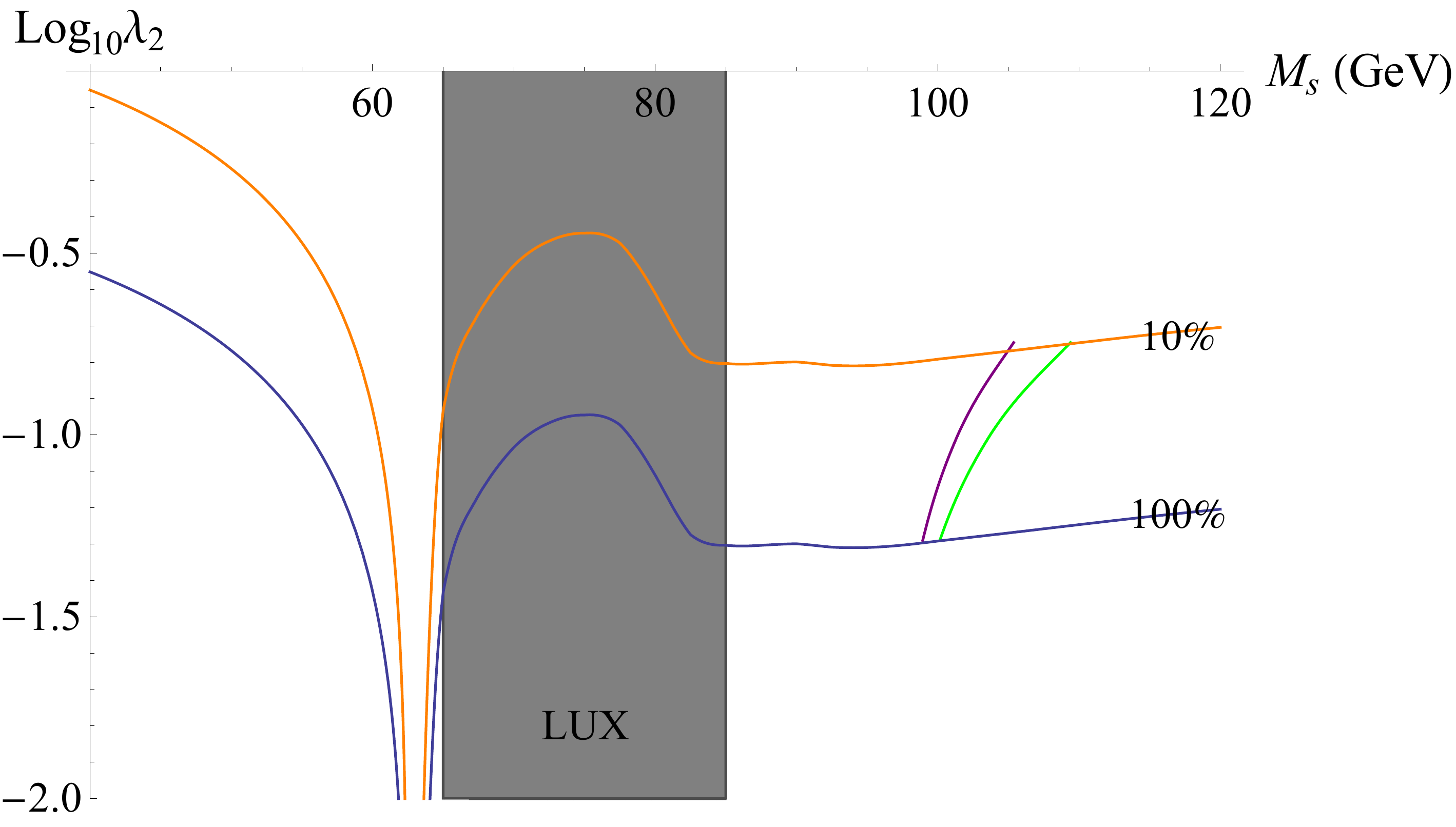}\hspace{0.05\columnwidth}
\caption{Relationship between predicted dark matter mass and Higgs portal coupling $\lambda_2$ with $\lambda_3=0$ is shown by the green curve and $\lambda_3=1$ shown by purple curve along with various dark matter abundance curves $10\%$ in yellow and $100\%$ in blue to constrain the complex singlet model. The shaded region represents the parameter space which is excluded by the LUX experiment at $95\%$ CL.}
\label{lux}
\end{figure}

We illustrate our predicted dark matter mass/coupling relation in the green curve in Fig.~\ref{lux}, which intersects the $10\%$ (orange) and $100\%$ (blue) dark matter abundance curves. These abundance curves are calculated using the results of Refs.~\cite{Steigman:2012nb,Cline:ab,Dittmaier:2011ti}. Compared to the real scalar model \cite{Steele:2013fka}, the complex singlet leads to a higher dark matter abundance because both components of the complex singlet contribute. Setting the dark matter self-interaction coupling to $\lambda_3=1$  shifts the results slightly from the green to the purple curve in the figure, retaining this qualitative feature.
The shaded region in Fig.~\ref{lux} represents the parameter space excluded by the LUX experiment at $95\%$ CL \cite{Akerib:2013tjd}, where we have 
followed the analysis of \cite{Cline:pe} and used the most conservative effective Higgs-nucleon coupling \cite{Rainer 2015} in the dark matter nucleon recoil cross section. Most of the parameter space below $85\,\rm{GeV}$ is ruled out by the LUX experiment \cite{Akerib:2013tjd}, apart from a small region of parameter space in the $M_S\approx M_{H}/2$ resonant region, which is strongly constrained by the Higgs decay width \cite{Djouadi:2012zc,Djouadi:2011aa} (see Refs.~\cite{Cline:ab,Cline:pe} for a comprehensive analysis). 
Combining the LUX \cite{Akerib:2013tjd} and dark matter abundance constraints, the complex singlet model admits a viable dark matter candidate $100\rm{GeV}\leq M_s\leq110\rm{GeV}$ with Higgs portal interaction $0.05\leq\lambda_2\leq0.2$ corresponding to $10\%-100\%$ dark matter abundance. 
The viable dark matter candidates resulting from our analysis are very close to the boundary of the current direct detection experiments and will be in the detection region of the coming experiments XENON1T \cite{XENON1T} and LUX 300 day results \cite{Akerib:2013tjd}.

We next consider the unbroken symmetry case with $y_S\neq 0$. The advantage of addressing diphoton excess in the unbroken case is twofold. First, in the unbroken case, there is no mixing between the singlet and the Higgs field, thereby
greatly suppressing the decay processes of the $S$ to  SM particles, in turn ideally satisfying the bounds  in ref.\cite{Franceschini:2015kwy}. Second, the $SHH$ term is forbidden by the global $U(1)$ symmetry; consequently the decay channels of the $S$ to SM particles through $SHH$ are prohibited, making this case an even better  candidate for  satisfying the bounds in \cite{Franceschini:2015kwy}.

We find when $\lambda_2\geq 3$, there exist two sets of coupling solutions. More interestingly, there occur two upper bounds for the singlet mass $M_S$, one for each set of the coupling solutions. For the first upper bound, we find $M_S\leq 217\,\rm{GeV}$ corresponding to $y_S\sim 0$. The maximal value of the Higgs portal coupling is $\lambda_2=5.6$; a larger Higgs portal coupling ($\lambda_2\geq 5.6$) will be non-perturbative and the above calculation method fails. Moreover, this upper bound will be sensitive to $y_S$, which decreases the bound. If we set $y_S\sim 1$ we find the upper bound decreases to $M_S\leq 133\,\rm{GeV}$.  
For the second upper bound, we find $M_S\leq 290\,\rm{GeV}$, corresponding to $y_S\sim 0$ and a Higgs portal coupling of $\lambda_2=4.5$. This upper bound is also be sensitive to $y_S$;  at $y_S\sim 1$ it decreases to $M_S\leq 247\,\rm{GeV}$. 

It is therefore almost impossible to push the singlet mass to $750\,\rm{GeV}$ for which the system possesses exact conformal symmetry with the global $U(1)$ symmetry unbroken. To address the diphoton excess, we will have to extend our model and introduce extra terms that provide extra contributions to the singlet mass (e.g.~another scalar portal couples to the singlet). A sample set of coupling solutions to address the $7\,\rm{fb}$ diphoton rate will be $y_S=1.35,~\lambda_1=2,~\lambda_3=0,~\lambda_2=2$ with the corresponding vector-like fermion mass at $M_F\sim830\,\rm{GeV}$ and the charge assignment $\left(R_C,R_W\right)_{Y_F}=\left(3,2\right)_{\frac{7}{6}}$. The predicted singlet mass is only around $90\,\rm{GeV}$;  an extra term is required to contribute the remaining $660\,\rm{GeV}$  in order to properly address the $750\,\rm{GeV}$ diphoton excess.

\section{Broken Phase}

The broken-symmetry case $\langle S\rangle\neq0$  is particularly interesting since the real component of the complex singlet will mix with the SM Higgs field, leading to one heavy and one light Higgs field. The light state corresponds to the $125\,\rm{GeV}$ observed Higgs boson \cite{2012gk,2012gu}  and the heavy state can potentially explain the recently observed $750\,\rm{GeV}$ diphoton resonance \cite{ATLAS,CMS} following the argument in \cite{Knapen:2015dap}. In this situation dark matter is associated with an axion decoupled from the effective potential, which in turn does not provide dark matter phenomenological constraints on the couplings present in the unbroken case. Consequently the guiding principles used to extract a meaningful range of the free parameter space of $\left(\lambda_2,\lambda_3\right)$ that remains after imposing Eqs.~\eqref{constraint2} are lost. To address this difficulty, we generalize our unbroken-symmetry methodology to incorporate a renormalization-scale optimization technique \cite{Casas:1994us}. This technique was used to obtain an optimized renormalization scale in the MSSM with conventional symmetry breaking (CSB) \cite{Casas:1994us}, and is based on the idea that the complete effective potential should be scale independent. Since we do not have  full information about the effective potential, which must be truncated at a particular loop order, the best that can be achieved is to find an optimized scale at which the scale-dependent minimum of the truncated effective potential self-consistently satisfies its RG equation. 

It is nontrivial to generalize this optimization method to incorporate  RSB. In the CSB scenario of the SM, all the couplings are known, and we only need to implement these known couplings as initial values and use the renormalization-group to run the couplings with the scale. The optimized scale is then explicitly determined by the point where the minimum of the effective potential satisfies its RG equation \cite{Casas:1994us}. However, in the case of RSB, all the couplings are unknown and should be determined dynamically from the theory itself \cite{Coleman:1973jx}. Without boundary values for the running couplings an intractable non-linear numerical problem occurs in determining the optimized scale.
To address this difficulty we  modify the optimization method to only depend on local quantities near the optimized scale, and define the scale-dependent minima $H_m(t)$ and $S_m(t)$ of the effective potential via 
\begin{gather}
F\left(H_m(t),S_m(t),t,\lambda_i(t)\right)=\frac{dV_{eff}}{dH}\bigg|_{H=H_m(t)\atop S=S_m(t)}=0\,, 
\\
G\left(H_m(t),S_m(t),t, \lambda_i(t)\right)=\frac{dV_{eff}}{dS}\bigg|_{{H=H_m(t)\atop S=S_m(t)}}=0\,,
\end{gather}
where $\mu=M_z\exp(t)$.  We then differentiate these constraints with respect to $t$, and impose the  condition \cite{Casas:1994us} 
\begin{equation}
\frac{dH_m(t^*)}{dt}=-\gamma(t^*)H_m(t^*)\,,~\frac{dS_m(t^*)}{dt}=0\label{global constraint}
\end{equation}
for the optimized scale $t*$, resulting in the two constraints
\begin{gather}
0=\frac{\partial F}{\partial t^*}-\gamma(t^*)H_m(t^*)\frac{\partial F}{\partial H_m}+\beta_i(t^*)\frac{\partial F}{\partial \lambda_i}\label{local constraint1}
\,,\\
0=\frac{\partial G}{\partial t^*}-\gamma(t^*)H_m(t^*)\frac{\partial G}{\partial H_m}+\beta_i(t^*)\frac{\partial G}{\partial \lambda_i}~\label{local constraint2}.
\end{gather}
Finally we connect the optimized minimum field configurations with the physical VEVs
\begin{equation}
H_m(t^*)=\langle H\rangle=v\,,~S_m(t^*)=v_1\,.
\end{equation}
Thus rather than requiring a global solution for $H_m(t)$ and $S_m(t)$ that is then used to determine $t^*$ via \eqref{global constraint}, we have encoded the same information into the local constraints \eqref{local constraint1}, \eqref{local constraint2} and the RG functions of the theory \cite{EliasMiro:2012ay}. 
Note that the CW renormalization condition \eqref{CW Condition} is unaffected except for the replacement $\mu=M_Z\exp(t^*)$.
It should be noted that in the case discussed in \cite{Casas:1994us}, only one optimization condition for the Higgs field is needed since they assumed supersymmetry is at a much higher scale and decoupled from the SM. We have generalized this optimization condition for both the Higgs and singlet fields, since the vacuum expectation value predicted here for the singlet may be near the electroweak scale, which cannot be decoupled. The above optimization conditions can be generalized further for more complicated models with multiple scalar fields.

Setting $y_S=0$ we have four constraints \eqref{constraint2}, \eqref{local constraint1}, \eqref{local constraint2} for five parameters $\lambda_1(t^*)$, $\lambda_2(t^*), \lambda_3(t^*), v_1, t^*$ where $\langle S\rangle=v_1$ is the VEV of the singlet field. Using the $125\,\rm{GeV}$ Higgs mass as an extra constraint, we find $\lambda_1(t^*)=0.53,~\lambda_3(t^*)=1.926,~\lambda_2(t^*)=-2.95,~\langle S\rangle=156\,\rm{GeV},~t^*=-1.59$ yielding  an additional  heavy Higgs at $554\,\rm{GeV}$.
The small scale $t^*=-1.59$ results from CW to MS scheme transformation $\mu_{CW}=\mu_{MS}/\lambda$ \cite{Steele:2014dsa,Ford:1991hw}, naturally leading to $\mu_{CW}\leq\mu_{MS}$.
We have also studied these couplings to assess their perturbative convergence using two loop RG functions \cite{Ballesteros:2015iua}. We found $\beta_1^{2 loop}/\beta_1^{1 loop}=5\times10^{-5},~\beta_2^{2 loop}/\beta_2^{1 loop}=0.04,~\beta_3^{2 loop}/\beta_3^{1 loop}=0.13$, which implies that higher-loop contributions are under control. Numerically similar Higgs portal couplings in two doublet models were found in Ref.~\cite{Hill:2014mqa}. The mixing angle is strongly constrained by the LHC and electroweak precision measurements \cite{Robens:2015gla} where LHC Higgs signal rates provide the strongest constraint $\sin\theta\leq 0.5$ in the region around a $500\,\rm{GeV}$ Higgs mass. In our model, we find a mixing angle $\sin\theta=0.467$, within the LHC run 2 detection region and not yet excluded. Note that higher loop effects might decrease the mixing angle further or alter the mass prediction of the heavier Higgs. 

We now set $y_S\neq 0$ and impose the constraints \eqref{constraint2}, \eqref{local constraint1}, \eqref{local constraint2}, requiring a $125\,\rm{GeV}$ Higgs mass and a second Higgs in the $750\,\rm{GeV}$ range. With these six constraints we find $y_S(t^*)=1.35,~\lambda_1(t^*)=1.73,~\lambda_3(t^*)=1.45,~\lambda_2(t^*)=-3.2,~\langle S\rangle=270\,\rm{GeV},~t^*=-0.55$ and the second Higgs to have mass $720\,\rm{GeV}$.
 Moreover, using the $7\,\rm{fb}$ fit value of the rate of the $750\,\rm{GeV}$ resonant production and decay to diphotons with the charge assignment $\left(R_C,R_W\right)_{Y_F}=\left(3,2\right)_{\frac{7}{6}}$ for the vector like fermion $F$ \cite{Knapen:2015dap}, we obtain a value of  $830\,\rm{GeV}$ for its mass $M_F$. The $830\,\rm{GeV}$ vector-like fermion mass satisfies the lower bounds $600\,\rm{GeV}$--$800\,\rm{GeV}$ provided in \cite{Aguilar-Saavedra:2013qpa}. Note that this value cannot be purely generated by the singlet fermion Yukawa term, since the $y_S$ Yukawa term only contributes $256\,\rm{GeV}$ to $M_F$ and a bare mass term is required. Thus, all the parameters in the system are determined. The mixing angle predicted in this case is $\sin\theta=0.67$, which satisfies the upper bounds ($\sin\theta\sim0.7$) of LHC SM Higgs searches and EW observables ($S, T, U$) for a second Higgs at $750\,\rm{GeV}$ provided in \cite{Robens:2015gla}. Further experimental results for the diphoton excess especially the searching of S decay channels to other SM particles will soon tell whether this scenario is viable \cite{Franceschini:2015kwy}. As a conclusion, our results of a $720\,\rm{GeV}$ second Higgs mass, $830\,\rm{GeV}$ vector-like fermion and a mixing angle of $\sin\theta=0.67$ are compatible with the current experimental bounds to address the $7\,\rm{fb}$ LHC diphoton excess. Note also that we have used the strongest version of the optimization method with Eq.~\eqref{constraint2}, \eqref{local constraint1}, \eqref{local constraint2}.
  
When the real component of the complex singlet obtains a VEV, the $U\left(1\right)$ global symmetry is spontaneously broken and generates a massless Goldstone boson containing the imaginary degree of freedom of the complex singlet. The complex singlet is conventionally written as $S\left(x\right)=\phi\left(x\right)\exp\left(\frac{ia\left(x\right)}{\sqrt2 f_a}\right)$ where $a\left(x\right)$ is the axion field and $f_a$ is the axion decay constant. 
Associating the $U\left(1\right)$ global symmetry with the Peccei-Quinn PQ symmetry \cite{Peccei:1977hh,Weinberg:1977ma}, the above Goldstone boson can be explained as the axion \cite{Meissner:2008gj, Latosinski:2010qm} which addresses the dark matter problem. Normally, a large intermediate scale is required to connect to the large PQ symmetry breaking scale to address the smallness of the axion coupling. However, any intermediate scale between the EW scale and UV scale is not allowed in the CW mechanism \cite{Bardeen,Shaposhnikov:2007nj}. In \cite{Meissner:2008gj, Latosinski:2010qm}, the authors cleverly connect the smallness of the axion coupling to the lightness of the neutrino mass and generate an effective large $f_a$ without introducing any large intermediate scale. Moreover, the $U\left(1\right)$ global symmetry considered in this work could also be made into a local symmetry, providing a new gauge interaction boson Z$^\prime$; symmetry breaking at the TeV scale  in this model was studied in Ref.\,\cite{Iso:2012jn}. 


It is interesting to analyze the underlying symmetry breaking mechanism for the broken case. We use the ratio of the tree-level VEV conditions as a measure of whether CSB or RSB is dominant. The ratio $r$ is defined by
\begin{equation}
r=\frac{dV_{tree}/d\phi^2}{dV_{tree}/d\varphi^2}\bigg|_{\phi=v\atop \varphi=v_1}=\frac{2\lambda_1\left(\frac{v^2}{v_1^2}\right)+\lambda_2}{2\lambda_3+\lambda_2\left(\frac{v^2}{v_1^2}\right)}
\end{equation}
where $r\ll1, r\gg1, r\simeq1$ correspond to CSB dominant in the Higgs sector, RSB dominant in the Higgs sector and the mixed scenario respectively. In the mixed scenario both CSB and RSB contribute to the EW symmetry breaking and we are not able to separate one from the other. Inputting the results $\lambda_1=0.53$, $\lambda_3=1.926, \lambda_2=-2.95, \langle S\rangle=156\,\rm{GeV}$ of the broken case, we obtain $r=0.1$ which implies  conventional EW symmetry breaking in the Higgs sector triggered by the CW mechanism in the hidden sector. Note that the Higgs quartic coupling $\lambda_1=0.53$ obtained in our case is around four times larger than the SM value of $\lambda_{\rm{SM}}=0.13$, implying comparatively large radiative corrections in the Higgs sector.

\section{Summary}

In summary, we have studied a conformally symmetric complex singlet extension of the SM with a Higgs portal interaction, whose global $U(1)$ symmetry is spontaneously broken or unbroken. The results have been summarized in Table \ref{summary}. In the unbroken case, radiative EW symmetry breaking in the SM Higgs sector is induced by the CW mechanism \cite{Coleman:1973jx}. The complex singlet is protected from decay, making it an ideal $\sim 100\,\rm{GeV}$ dark matter candidate comprising a
 significant proportion of the thermal relic abundance that is within the detection region of the upcoming XENON1T \cite{XENON1T} and LUX 300 day \cite{Akerib:2013tjd} experiments.  Including an extra vector-like fermion $F$, this case can also provide an ideal explanation for diphoton excesses without violating the experimental bounds only if extra terms are introduced to increase the singlet mass. In the broken case,  generalizing and improving upon the optimization method inspired by \cite{Casas:1994us}, we found a sequential symmetry breaking scenario, in which RSB in the singlet sector triggers  conventional EW symmetry breaking in the Higgs sector. We found there exists a second Higgs boson with an approximate $550\,\rm{GeV}$ mass and a mixing angle $\sin\theta\approx0.47$, which satisfies the current experiment bound $\sin\theta\leq 0.5$ at around the $500\,\rm{GeV}$ Higgs mass region provided by the LHC signal rates \cite{Robens:2015gla} that will be strengthened during LHC run 2. Moreover, including the extra vector-like fermion $F$ we find a set of coupling solutions where the  second Higgs boson mass increases to around $720\,\rm{GeV}$ and the extra vector-like fermion mass is $830\,\rm{GeV}$,  addressing the $750\,\rm{GeV}$ diphoton anomaly observed at the LHC \cite{ATLAS,CMS}.

\begin{table}[ht]
\centering
  \begin{tabular}{|| l | l | l | l | l ||}
    	\hline
Scenarios & Dark Matter & Diphoton Excess & Second Higgs & $\sin\theta$ \\ \hline Unbroken; $y_S=0$ & Yes; Cold & No & No & $0$ \\ \hline Unbroken; $y_S\neq 0$ & No & No; Singlet mass too small & No & $0$ \\ \hline Broken; $y_S=0$ & Yes; Axion & No & Yes; $550\,\rm{GeV}$ & $0.47$ \\ \hline Broken; $y_S\neq 0$ & No & Yes & Yes; $720\,\rm{GeV}$ & $0.67$ \\ \hline
\end{tabular}
\caption{Two categories (unbroken and broken phase) and four scenarios (each phase with either $y_S=0$ or $y_S\neq 0$ where $y_S$ is the scalar-vector like fermion coupling) are summarized in the table where $\sin\theta$ corresponds to the mixing angle between the Higgs field and the singlet.}
\label{summary}
\end{table}

\acknowledgments

T.G.S. and R.B.M are grateful for financial support from the Natural Sciences and Engineering Research Council of Canada (NSERC).



\end{document}